\documentclass[prl,twocolumn,showpacs,superscriptaddress]{revtex4}
\usepackage{graphicx}
\usepackage{indentfirst}
\usepackage{epsfig}
\usepackage{subfigure}
\usepackage{amsmath}
\usepackage{amssymb}
\def\be{\begin{eqnarray}}
\def\ee{\end{eqnarray}}

\begin{document}
\title{Dipole emission and coherent transport in random media III.\\
      Emission from a real cavity in a continuous medium}
\author{M. Donaire}
\email{manuel.donaire@uam.es}
\address{Departamento de F\'{\i}sica de la Materia Condensada, Universidad Aut\'{o}noma de
Madrid, E-28049 Madrid, Spain.}
\begin{abstract}
This is the third of a series of papers devoted to develop a
microscopical approach to the  dipole emission process and its
relation to coherent transport in random media. In this paper, we
compute the power emitted by an induced dipole and the spontaneous
decay rate of a Lorentzian-type dipole. In both cases, the emitter
is placed at the center of a real cavity drilled in a continuous
medium.
\end{abstract}
%
 \maketitle
\indent It is known that the spontaneous emission rate, $\Gamma$,
 in a dielectric medium depends on the interaction of the emitter with the environment \cite{Purcell}. This is so because the
 surrounding medium determines the number of channels through which the excited particle can emit. That is, the local
 density of states (LDOS). For practical purposes the understanding of the spontaneous emission rate of a fluorescence particle is of great interest in biological imaging \cite{Suhling}. Closely related is the process of emission/reception by  nano-antennas \cite{Antenas}. Also, inhibition of spontaneous emission is expected to occur in photonic band gap materials \cite{JohnI} as a signature of localization.\\
\begin{figure}[h]
\includegraphics[height=3.4cm,width=6.7cm,clip]{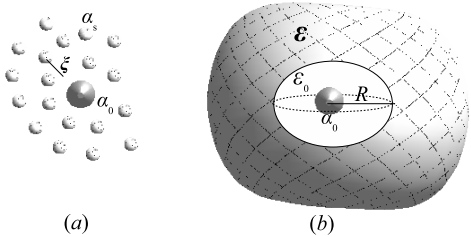}
\caption{($a$) Cermet topology scenario. The emitter with polarizability $\alpha_{0}$ is
 surrounded by a disconnected net of scatterers of polarizability $\alpha_{s}$ and correlation length $\xi$. ($b$) Simply-connected-non-contractible topology in which the emitter is placed at the center of a cavity of radius $R$ surrounded by a continuous medium of dielectric contrast $\epsilon$.}\label{fig32}
\end{figure}
 \indent In a previous paper \cite{MeI}, general analytical formulae were derived for LDOS and $\Gamma$ as functions of both the electrical susceptibility of the medium and the geometry of the embedding of the emitter in it. There, three different cases are addressed attending to the nature of the emitter. In the first case, the emitter is a polarizable dipole of transition amplitude $\mu$ at resonance. In the second case, the emitter is a non-resonant dipole of polarizability $\alpha$ with induced dipole moment $\epsilon_{0}\alpha\vec{E}_{0}$, where $\vec{E}_{0}$ is a fixed exciting field. In the third case, the emitter is a fluorescent particle seated on top of a polarizable host particle. Attending to the embedding of the emitter and the topology of the system emitter-host-medium, we differentiated two cases \cite{MeI}. In the first one, the emitter is equivalent, replaces or is placed on top of a host scatterer. In such a
case, by consistency, the topology of the medium is that of a disconnected manifold --known also as \emph{cermet
topology}-- in the sense that the medium is made of spherical
inclusions and the emitter takes the place of one of those
inclusions. Attending to the embedding, the emitter is placed within a virtual cavity but for the case it replaces a host scatterer. The spatial correlation of the emitter with the
surrounding medium is given by the same correlation
function which enters the electrical susceptibility tensor,
$\bar{\chi}$. Spatial dispersion in $\bar{\chi}$ is inherent to
that topology. This scenario was studied in \cite{MeII} with an emitter seated on top of a host scatterer. A
relation between the complex refraction index of the host
medium and the spontaneous decay rate was found.\\
\indent In this paper we will focus on the remaining topological case.
 The emitter is seated at the center of a spherical cavity
drilled in a homogeneous medium. This medium is continuous as seen by the
emitter provided that the cavity radius, $R$, is much larger than any correlation length between the
 constituents of the host medium, $\xi$. Thus, the cavity is real. On the other hand, if the wavelength of the emitted field, $\lambda$, is that long that it cannot resolve neither the microscopical structure of the medium nor the cavity, the medium is also continuous as seen by the propagating field. Consequently, the host medium is characterized by a scalar
susceptibility $\chi$ which does not present spatial
dispersion. Topologically, this setup corresponds to that of a
 simply-connected-non-contractible manifold with a unique hollow sphere --see Fig.\ref{fig32}. Regarding the emitter nature, two scenarios will be addressed for their potential interest in biological imaging and signal processing. In the first one, the emitter will be a polarizable particle excited by an external fixed field of frequency much lower than any of the resonance frequencies of the emitter. The physical quantity to compute in this case will be the power emission, $W^{\alpha}$. In the second case, the emitter will be a polarizable particle which decays at some resonance frequency. Dipole induction will be encoded in the emitter itself as it will be assumed to be of Lorentz-type. Thus, the net effect of the host medium will reflect on a variation in the spontaneous decay rate, $\Gamma^{\alpha}$, together with a shift in the resonance frequency with respect to those values in free space.\\
\indent In the first scenario, the formula for the power emitted by an induced dipole reads
\begin{eqnarray}
W^{\alpha}&=&
\frac{\omega_{0}\epsilon_{0}^{2}}{2}\Im{}\Bigl\{\frac{\alpha_{0}}{1+\frac{1}{3}k_{0}^{2}\alpha_{0}[2\gamma_{\perp}+\gamma_{\parallel}]}
\Bigr\}|E^{\omega}_{0}|^{2}\label{la1}\\&=&\frac{-\omega_{0}^{3}\epsilon_{0}^{2}}{6c^{2}}\Bigl[\frac{|\alpha_{0}|^{2}}
{|1+\frac{1}{3}k_{0}^{2}\alpha_{0}[2\gamma_{\perp}+\gamma_{\parallel}]|^{2}}\Im{\{2\gamma_{\perp}+\gamma_{\parallel}\}}\label{la2}\\
&-&
\frac{\Im{\{\alpha_{0}\}}}{|1+\frac{1}{3}k_{0}^{2}\alpha_{0}[2\gamma_{\perp}+\gamma_{\parallel}]|^{2}}\Bigr]|E^{\omega}_{0}|^{2}\label{la3},
\end{eqnarray}
where $\omega_{0}$ is the frequency of the exciting field $\vec{E}^{\omega}_{0}$, with $k_{0}=\omega_{0}/c$ much lower than any internal resonance frequency and $\alpha_{0}\equiv 4\pi a^{3}\frac{\epsilon_{e}-1}{\epsilon_{e}+2}$ is the electrostatic polarizability of the emitter in vacuum,
being $\epsilon_{e}(\omega_{0})$ its dielectric contrast and $a$ its radius.
For our topological setup, the $\gamma$-factors
$\gamma_{\perp,\parallel}$ are given by \cite{MeI}
\begin{eqnarray}
2\gamma_{\perp}(\tilde{k})&=&-i\frac{\tilde{k}}{2\pi}-2\tilde{k}^{2}\int\frac{\textrm{d}^{3}k}{(2\pi)^3}
[C_{\perp}+G_{\perp}^{(0)}]\chi_{\perp}G_{\perp}^{(0)}(k)\nonumber\\&+&
2\tilde{k}^{4}\int\frac{\textrm{d}^{3}k}{(2\pi)^3}[C_{\perp}+G^{(0)}_{\perp}]^{2}
G_{\perp}\chi_{\perp}^{2}(k),\label{OB2a}\\
\gamma_{\parallel}(\tilde{k})&=&-\tilde{k}^{2}\int\frac{\textrm{d}^{3}k}{(2\pi)^3}
[C_{\parallel}+G_{\parallel}^{(0)}]\chi_{\parallel}G_{\parallel}^{(0)}(k)\nonumber\\&+&
\tilde{k}^{4}\int\frac{\textrm{d}^{3}k}{(2\pi)^3}[C_{\parallel}+G^{(0)}_{\parallel}]^{2}
G_{\parallel}\chi_{\parallel}^{2}(k),\label{OB2b}
\end{eqnarray}
with $\tilde{k}$ evaluated at $k_{0}$ and
where the cavity factors $C_{\perp,\parallel}$ are
\begin{figure}[h]
\includegraphics[height=6.2cm,width=8.4cm,clip]{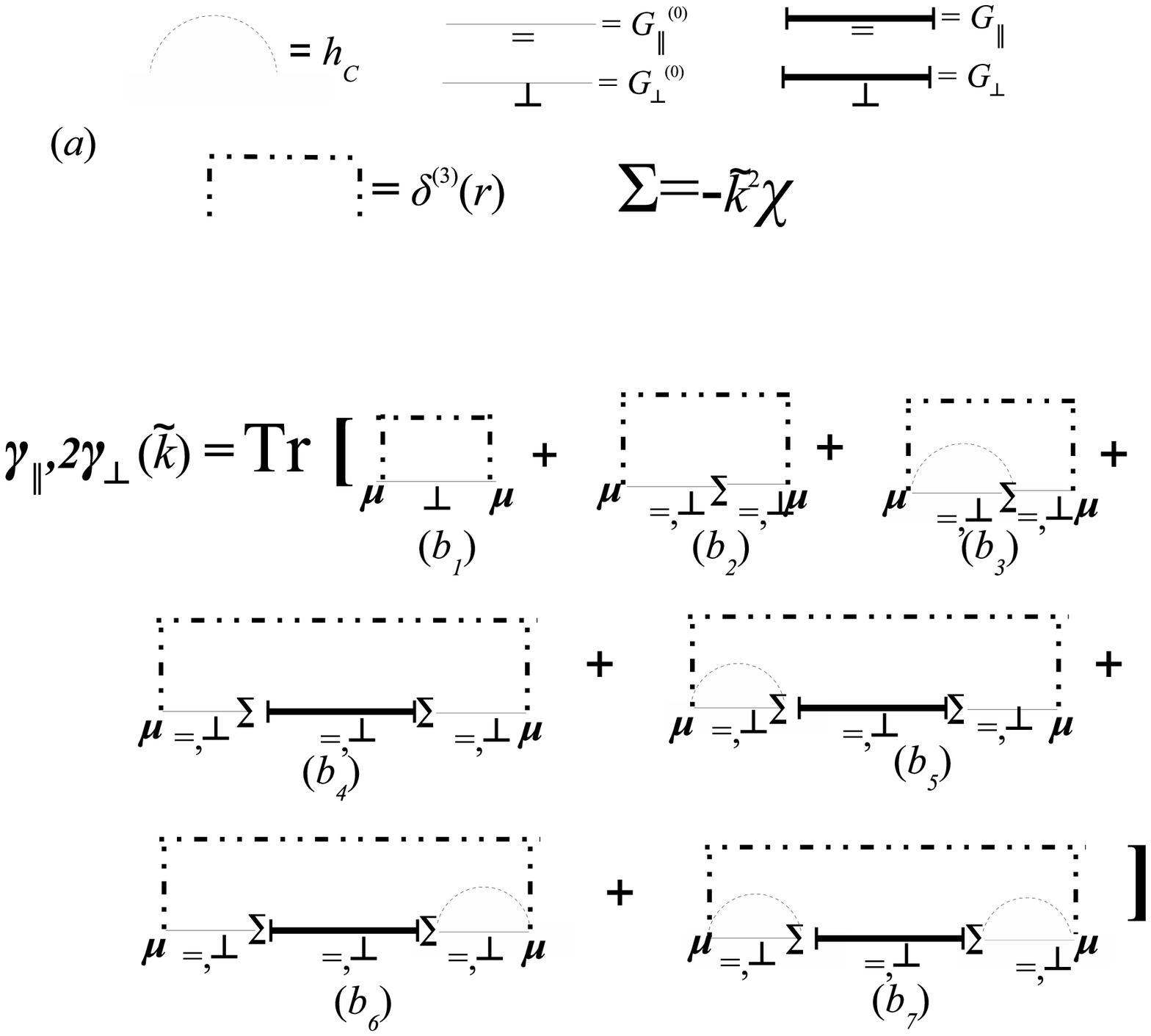}
\caption{($a$) Feynman's rules . ($b$) Diagrammatic representation
of the transverse and longitudinal polarization factors,
$2\gamma_{\perp}$ and $\gamma_{\parallel}$, as given by
Eqs.(\ref{OB2a},\ref{OB2b}).}\label{fig31}
\end{figure}
\begin{eqnarray}
C_{\perp}(k)&=&\frac{1}{2}\int\textrm{d}^{3}r\:e^{i\vec{k}\cdot\vec{r}}h_{C}(r)\textrm{Tr}
\{\bar{G}^{(0)}(r)[\bar{I}-\hat{k}\otimes\hat{k}]\}\nonumber\\&=&
\frac{1}{2}\int\frac{\textrm{d}^{3}k'}{(2\pi)^{3}}
h_{C}(|\vec{k}'-\vec{k}|)\Bigl[G_{\perp}^{(0)}(k')\nonumber\\&+&G_{\perp}^{(0)}(k')\cos^{2}{\theta}\:+\:G_{\parallel}^{(0)}(k')\sin^{2}{\theta}\Bigr],
\label{Xiperp}\\
C_{\parallel}(k)&=&\int\textrm{d}^{3}r\:e^{i\vec{k}\cdot\vec{r}}h_{C}(r)\textrm{Tr}
\{\bar{G}^{(0)}(r)[\hat{k}\otimes\hat{k}]\}\nonumber\\&=&\int\frac{\textrm{d}^{3}k'}{(2\pi)^{3}}
h_{C}(|\vec{k}'-\vec{k}|)\nonumber\\&\times&\Bigl[G_{\parallel}^{(0)}(k')\cos^{2}{\theta}\:+\:G_{\perp}^{(0)}(k')\sin^{2}{\theta}\Bigr],
\label{Xiparall}
\end{eqnarray}
with $\cos{\theta}\equiv \hat{k}\cdot\hat{k}'$.
In these equations, $h_{C}(k)$ is the Fourier transform of the correlation function which describes the isolation of the emitter within a real cavity.
 For a spherical cavity of radius $R$, $h_{C}(r)=-\Theta(r-R)$.  In the above equations $G_{\perp}$ and $G_{\parallel}$ are respectively the
 transverse and longitudinal components of the renormalized propagator
 of the coherent --macroscopic-- electric field in the host
 medium. They read,
 $\bar{G}_{\perp}(k)=\frac{\Delta(\hat{k})}{\tilde{k}^{2}\epsilon_{\perp}(k)-k^{2}}$,
$\bar{G}_{\parallel}(k)=\frac{\hat{k}\otimes\hat{k}}{\tilde{k}^{2}\epsilon_{\parallel}(k)}$,
$\hat{k}$ being a unitary vector along the propagation direction
and $\Delta(\hat{k})\equiv I-\hat{k}\otimes\hat{k}$ being the
projective tensor orthogonal to the propagation direction. In
these expressions,
$\epsilon_{\perp,\parallel}(k)=1+\chi_{\perp,\parallel}(k)$.
Because no spatial dispersion is assumed in the continuous medium,
$\epsilon_{\perp}=\epsilon_{\parallel}\equiv\epsilon$ are functions
 of $\tilde{k}$ only. $\bar{G}^{(0)}_{\perp,\parallel}$ are the
components of the propagator of the electric field in free space with bare wave number $\tilde{k}$.
They are given by the above formulae, $\bar{G}_{\perp}$, $\bar{G}_{\parallel}$, with
$\epsilon_{\perp}=\epsilon_{\parallel}=1$.\\
\indent In the second scenario, the emitter nature differs from that in \cite{MeII}. In the present case, the dipole moment of the emitter is not a linear combination of a fixed and an induced dipole. Rather than that, the emitter is polarizable in its own and its polarizability function $\alpha(\tilde{k})$ gets renormalized as a
result of iterative self-polarization processes,
\begin{equation}\label{Lorentzpol}
\alpha(\tilde{k})=\alpha_{0}[1+\frac{1}{3}\alpha_{0}\tilde{k}^{2}\Re{\{2\gamma_{\perp}^{(0)}\}}
+\frac{1}{3}\alpha_{0}\tilde{k}^{2}(2\gamma_{\perp}+\gamma_{\parallel})]^{-1}.
\end{equation}
In the above equation, the $\gamma$-factors are those in Eqs.(\ref{OB2a},\ref{OB2b}). In addition, the term
$\frac{1}{3}\alpha_{0}\tilde{k}^{2}\Re{\{2\gamma_{\perp}^{(0)}\}}$ has been introduced to account for the internal resonance
of the emitter in vacuum. As shown in \cite{deVriesRMP}, it plays the role of a regulator of the intrinsic ultraviolet divergence in $G^{(0)}_{\perp}$. That is, $\Re{\{2\gamma_{\perp}^{(0)}\}}=\frac{-3}{k_{0}^{2}\alpha_{0}}$, where $\alpha_{0}$ is the corresponding real electrostatic polarizability and $k_{0}$ is the resonance wave number in vacuum. Following \cite{deVriesPRL}, by parametrizing
Eq.(\ref{Lorentzpol}) in  a Lorentzian (L) form, $\alpha_{L}(\tilde{k})=\alpha'_{0}k_{res}^{2}[k_{res}^{2}-\tilde{k}^{2}-i\Gamma^{\alpha}\tilde{k}^{3}/(c k_{res}^{2})]^{-1}$,
we can identify the decay rate of the emitter in the host medium as
\begin{eqnarray}\label{Gamares}
\Gamma^{\alpha}&=&-\frac{c}{3}\alpha'_{0}\tilde{k}^{3}\Im{\{2\gamma_{\perp}+\gamma_{\parallel}\}}|_{\tilde{k}=k_{res}}
\nonumber\\&=&-\Gamma_{0}\frac{2\pi}{k_{0}^{2}}\tilde{k}\Im{\{2\gamma_{\perp}+\gamma_{\parallel}\}}|_{\tilde{k}=k_{res}},
\end{eqnarray}
where $k_{res}$ is a real non-negative root of the equation
\begin{equation}\label{kres}
(\tilde{k}/k_{0})^{2}-1=\frac{1}{3}\alpha_{0}\tilde{k}^{2}\Re{\{2\gamma_{\perp}+\gamma_{\parallel}\}}|_{\tilde{k}=k_{res}},
\end{equation}
$\alpha'_{0}=\alpha_{0}(k_{0}/k_{res})^{2}$ is the renormalized electrostatic polarizability and $\Gamma_{0}=c\alpha_{0}k_{0}^{4}/6\pi$ is the in-vacuum emission rate. As argued in \cite{deVriesPRL}, consistency with Fermi's golden
rule requires $\alpha_{0}=\frac{2|\mu|^{2}}{\epsilon_{0}\hbar c k_{0}}$, $\mu$ being the transition amplitude between
two atomic levels in vacuum.\\
\indent In view of Eqs.(\ref{la1},\ref{la2},\ref{la3},\ref{Gamares}), we can address in parallel the computation of $W^{\alpha}$ and $\Gamma^{\alpha}$ for each
scenario described above.  The computation reduces to the evaluation of $2\gamma_{\perp}(\tilde{k})+\gamma_{\parallel}(\tilde{k})$ for $\tilde{k}=k_{0}$ and $\tilde{k}=k_{res}$ respectively.\\
\indent Following \cite{MeI,MeII}, we will decompose $\Gamma^{\alpha}$ and $W^{\alpha}$ in
transverse and longitudinal components, $\Gamma^{\alpha}=2\Gamma^{\alpha}_{\perp}+\Gamma^{\alpha}_{\parallel}$ and
$W^{\alpha}=2W^{\alpha}_{\perp}+W^{\alpha}_{\parallel}$ respectively.
 Neglecting longitudinal resonances as in
\cite{MeII}, only the transverse components contain propagating
modes, $\Gamma_{P}^{\alpha}$ and $W_{P}^{\alpha}$ respectively,
which are proportional to the coherent intensity and so to
$\Im{\{2\gamma^{P}_{\perp}(\tilde{k})\}}$. It is convenient to
identify their specific contribution to $\Gamma^{\alpha}$ and
$W^{\alpha}$  for observational purposes in imaging. Propagating
modes are those associated to complex poles of the propagators in
the integrands of eq.(\ref{OB2a}). We can read their contribution
from the diagrams in Fig.\ref{fig31},
\begin{eqnarray}
\Im{\{2\gamma_{\perp}^{P}(\tilde{k})\}}=2\int\Im{\{G_{\perp}(k)\}}\frac{\textrm{d}^{3}k}{(2\pi)^3}\nonumber\\
-2\tilde{k}^{2}\Re{\{\chi
C_{\perp}(\tilde{k})\}}\int\Im{\{G^{(0)}_{\perp}(k)\}}\frac{\textrm{d}^{3}k}{(2\pi)^3}\nonumber\\
+2\tilde{k}^{4}\Re{\{[\chi
C_{\perp}(\tilde{k})]^{2}\}}\int\Im{\{G_{\perp}(k)\}}\frac{\textrm{d}^{3}k}{(2\pi)^3}\nonumber\\
+4\tilde{k}^{4}\Re{\{\chi
C_{\perp}(\tilde{k})\}}\int\Im{\{G_{\perp}^{(0)}(k)\chi
G_{\perp}(k)\}}\frac{\textrm{d}^{3}k}{(2\pi)^3},\label{firstp}
\end{eqnarray}
where the first term on the \emph{right hand side} (\emph{r.h.s}) of Eq.(\ref{firstp}) contains the contribution of
bulk propagation --transverse components of diagrams ($b_{1}$),
($b_{2}$), ($b_{4}$) in Fig.\ref{fig31}.\\
\indent All the above integrals in Eqs.(\ref{OB2a}-\ref{Xiparall},\ref{firstp}) can be performed analytically at any order in $\tilde{k}R$. However, for the sake of consistency with the approximations considered so far, we take the limit
$\tilde{k}R\ll1$ and keep leading order terms,
\begin{eqnarray}\label{gamis}
2\gamma_{\perp}+\gamma_{\parallel}\simeq-\frac{i}{2\pi}\tilde{k}\Bigl[\Bigl(\frac{\epsilon+2}{3}\Bigr)
^{2}\sqrt{\epsilon}-\frac{4}{9}\frac{(\epsilon-1)^{2}}{\epsilon}\Bigr]\nonumber\\
-\frac{\tilde{k}}{2\pi}\Bigl[\Bigl(\frac{1}{(\tilde{k}R)^{3}}+\frac{1}{\tilde{k}R}\Bigr)
[(\epsilon-1)\frac{\epsilon+2}{3\epsilon}]\Bigr].
\end{eqnarray}
\indent It is our choice to normalize $W^{\alpha}$ to the propagating power in vacuum,
$W_{0}=\frac{\omega_{0}^{4}\epsilon_{0}^{2}}{12\pi c^{3}}\frac{|\alpha_{0}|^{2}}
{|1-\frac{i}{6\pi}k_{0}^{3}\alpha_{0}|^{2}}|E^{\omega}_{0}|^{2}$. Thus, $W^{\alpha}$ reads
\begin{eqnarray}
W^{\alpha}&=&W_{0}\mathcal{R}_{p}(\alpha_{0})\Bigl[\Re{\{\Bigl(\frac{\epsilon+2}{3}\Bigr)^{2}\sqrt{\epsilon}\}}-\frac{4}{9}\Re{\{\frac{(\epsilon-1)^{2}}{\epsilon}\}}
\label{firstyp}\\
&-&\Bigl(\frac{1}{(k_{0}R)^{3}}+\frac{1}{k_{0}R}\Bigr)\Im{\{\frac{2(\epsilon-1)^{2}}{3\epsilon}-(\epsilon-1)\}}
\label{secyp}\\&+&\Im{\{\alpha_{0}\}}\Bigr]\label{thirdyp},
\end{eqnarray}
where $\mathcal{R}_{p}(\alpha_{0})=|1+\frac{1}{3}k_{0}^{2}\alpha_{0}[2\gamma_{\perp}(k_{0})+\gamma_{\parallel}(k_{0})]|^{-2}$
is the renormalization factor due to self-polarization cycles.
Written this way, we recognize in the first term of the \emph{r.h.s} of Eq.(\ref{firstyp}) the usual
bulk term corrected by Lorentz-Lorenz (LL) local field factors
\cite{Lorentz},
$\Gamma^{LL}=\Gamma_{0}\Re{\{\Bigl(\frac{\epsilon+2}{3}\Bigr)^{2}\sqrt{\epsilon}\}}$.
The second term there includes corrections of order
$\gtrsim\chi^{2}$. It is also remarkable that if the empty-cavity Onsager--B\"{o}ttcher (OB) \cite{Onsager,GL} local field factors are used instead, $\Gamma^{OB}=\Gamma_{0}\Re{\{\Bigl(\frac{3\epsilon}{2\epsilon+1}\Bigr)^{2}\sqrt{\epsilon}\}}$
 does agree with the two terms of Eq.(\ref{firstyp}) up to order $\chi^{2}$. The terms in Eq.(\ref{secyp}) are
associated to absorbtion in the host medium \cite{Loudon,MeII} while that in Eq.(\ref{thirdyp}) corresponds to absorbtion in the emitter. The propagating emission reads
\begin{eqnarray}\label{propar}
W_{P}^{\alpha}&=&W_{0}\mathcal{R}_{p}(\alpha_{0})\Im{\{2\gamma^{P}_{\perp}(k_{0})\}}\nonumber\\
&=&W_{0}\mathcal{R}_{p}(\alpha_{0})\Bigl[\Re{\{\sqrt{\epsilon}\}}\Re{\Bigl\{\Bigl(\frac{\epsilon+2}{3}\Bigr)^{2}\Bigr\}}
\nonumber\\
&-&\frac{1}{3}\Re{\{\epsilon-1\}}\Bigr].
\end{eqnarray}
As shown in \cite{MeII} at leading order, the second term in
Eq.(\ref{propar}) subtracts from the total $\Gamma^{LL}$ the
non-propagating, non-absorptive longitudinal contribution.\\
 \indent It is worth comparing this result with that obtained in \cite{MeII}.
 There, provided that $\epsilon$ follows Maxwell-Garnett (MG)  formula  \cite{vanTiggelen} and neglecting absorbtion,
\begin{equation}\label{theother}
W^{MG}_{P}\simeq W_{0}\mathcal{R}^{MG}_{p}(\alpha_{0})\Bigl[\Re{\{\sqrt{\epsilon}\}}\Re{\{\frac{\epsilon+2}{3}\}}\Bigr].
\end{equation}
Eq.(\ref{propar}) and Eq.(\ref{theother}) agree at leading order
in $\chi$. However, in \cite{MeII} the cermet topology was
considered and the emitter was placed at the site of one of the
host scatterers.
The coincidence of both approaches at leading order bases on the fact that the limit $k_{0}\xi\ll1$, with $\xi$ being the radius of the exclusion volume between host scatterers, is implicit in Eq.(\ref{theother}). Therefore, no spatial  dispersion enters $\chi$ --hence, MG holds-- and the radius of the virtual cavity $R=\xi$ also satisfies $k_{0}R\ll1$ as in the present case.\\
\indent It is straightforward to write $W_{P}^{\alpha}$ in function of
the effective refraction index $\bar{n}$ and the extinction coefficient $\kappa\equiv (2k_{0}l_{ext})^{-1}$, with $l_{ext}$ the extinction mean
free path, using the identities
$\Re{\{\chi\}}=\bar{n}^{2}-1-\kappa^{2}$
and $\Im{\{\chi\}}=2\bar{n}\kappa$,
\begin{eqnarray}
W_{P}^{\alpha}&=&W_{0}\mathcal{R}_{p}(\alpha_{0})\Bigl[\frac{\bar{n}}{9}(4+\bar{n}^{4}+\kappa^{4}-6\bar{n}^{2}\kappa^{2}
+4\bar{n}^{2}-4\kappa^{2})\nonumber\\
&-&\frac{1}{3}(\bar{n}^{2}-1-\kappa^{2})\Bigr]\label{n2}.
\end{eqnarray}
For the sake of completeness, we give also the values of the non-propagating emission in terms of $\bar{n}$, $\kappa$ and $\alpha_{0}$,
\begin{eqnarray}
W_{NP}^{\alpha}&=&W_{0}\mathcal{R}_{p}(\alpha_{0})\Bigl\{\Im{\{\alpha_{0}\}}\label{abs1}\\
&+&\Bigl[\frac{2\kappa^{2}\bar{n}}{9}(1-\frac{4}{(\kappa^{2}+\bar{n})^{2}})\Bigr]\label{abs2}\\
&+&\frac{2\bar{n}\kappa}{3}\Bigl(\frac{1}{(k_{0}\xi)^{3}}+\frac{1}{k_{0}\xi}\Bigr)
\Bigl(1+\frac{2}{(\bar{n}^{2}+\kappa^{2})^{2}}\Bigr)
\nonumber\\&+&\frac{5}{9}\Bigl(1+\frac{\kappa^{6}-\bar{n}^{6}+(\kappa^{2}-\bar{n}^{2})(4-\kappa^{2}\bar{n}^{2})}
{5(\kappa^{2}+\bar{n}^{2})^{2}}\Bigr)\Bigr\}\label{abs3},
\end{eqnarray}
where Eqs.(\ref{abs1},\ref{abs2},\ref{abs3}) correspond to the power absorbed by the emitter, the transverse non-propagating power, $W^{NP}_{\perp}$, and the longitudinal emission, $W^{\alpha}_{\parallel}$, respectively.
Further on, by invoking causality, Kramers-Kronig sum rules relate $\bar{n}$ and
$l_{ext}$ and variations of
$W^{\alpha}$  can be obtained as a function of $\omega_{0}$ \cite{LoudonII}.\\
\indent Let us introduce next the self-polarization effect. This is done by multiplying all the expressions above for $W^{\alpha}$ by the renormalization factor $\mathcal{R}_{p}(\alpha_{0})$. Neglecting absorbtion  both in the emitter and in the host medium, we obtain at leading order in $k_{0}R$,
\begin{equation}\label{renormy}
\mathcal{R}_{p}(\alpha_{0})\simeq\Bigl(1-\frac{2}{3}\frac{\alpha_{0}}{V_{R}}\frac{\epsilon-1}{3\epsilon}
\frac{\epsilon+2}{3}\Bigr)^{-2},
\end{equation}
where $V_{R}$ is the volume of the emitter cavity. In contrast, the OB renormalization factor due to self-polarization reads \cite{Onsager,Bullough,deVriesPRL},
\begin{equation}\label{renormOB}
\mathcal{R}_{p}^{OB}(\alpha_{0})=\Bigl(1-\frac{2}{3}\frac{\alpha_{0}}{V_{R}}
\frac{\epsilon-1}{3\epsilon}
\frac{3\epsilon}{2\epsilon+1}\Bigr)^{-2}.
\end{equation}
Eq.(\ref{renormy}) and Eq.(\ref{renormOB}) differ just in the local field factor which multiplies
$-\frac{2}{3}\frac{\alpha_{0}}{V_{R}}\frac{\epsilon-1}{3\epsilon}$ in each expression. With this result we finalize the computation of $W^{\alpha}$ in the first scenario.\\
\indent In the following, we address the scenario where the emitter is of Lorentzian kind and compute
Eqs.(\ref{Lorentzpol}-\ref{kres}). We first solve for the renormalized resonance wave vector $k_{res}$,
\begin{eqnarray}
k_{res}&\simeq& k_{0}\Bigl[1-\frac{\alpha_{0}}{9V_{R}}\Re{\{(\epsilon-1)\frac{\epsilon+2}{3\epsilon}\}}\label{kres1}\\
&+&\frac{\alpha_{0}k_{0}^{3}}{6\pi}\Im{\{\Bigl(\frac{\epsilon+2}{3}\Bigr)
^{2}\sqrt{\epsilon}-\frac{4}{9}\frac{(\epsilon-1)^{2}}{\epsilon}\}}\Bigr]\label{kres2}.
\end{eqnarray}
At leading order in $k_{0}R$, the first two terms in Eq.(\ref{kres1}) equal the expression found in \cite{deVriesPRL}
for an interstitial emitter. The additional terms in Eq.(\ref{kres2}) are relevant for
strongly absorptive media. That may be for instance the case of nano-antennas in metallic media \cite{Antenas}. Additional solutions  to the one above for $k_{res}$ are only admissible beyond
the small cavity limit and/or the continuous medium approximation. That is, for
$\tilde{k}R\gtrsim1$ and/or $\tilde{k}\xi\gtrsim1$, Eq.(\ref{kres}) may contain  several real solutions. That
might give rise to a splitting of the in-vacuum resonant peak provided such an splitting is broader than
the line width. The origin of the resonance shifts of Eqs.(\ref{kres1},\ref{kres2}) is in the virtual photons
which dress up the polarizability $\alpha$ through Eq.(\ref{Lorentzpol}). Those photons trace closed
spatial loops as depicted by the diagrams in Fig.(\ref{fig31}). They amount to a classical continuum of
sates not to be confused with real localized photons. The resultant frequency shift must be rather
interpreted as the classical analog to the quantum Lamb shift \cite{Wylie,Sakurai}. On the other hand, the possibility  of a
 resonance splitting would be subjected to
consistency with our perturbative renormalization scheme and linear-response approximation.\\
\indent Finally, we give the full expression for $\Gamma^{\alpha}$ at lowest order in $k_{0}R$ according to Eq.(\ref{Gamares}),
\begin{eqnarray}\label{Gamores}
\Gamma^{\alpha}&\simeq&\Gamma_{0}\Bigl[[1-\frac{2\alpha_{0}}{9V_{R}}\Re{\{(\epsilon-1)\frac{\epsilon+2}{3\epsilon}\}}\nonumber\\
&+&\frac{2\alpha_{0}k_{0}^{3}}{6\pi}\Im{\{\Bigl(\frac{\epsilon+2}{3}\Bigr)
^{2}\sqrt{\epsilon}-\frac{4}{9}\frac{(\epsilon-1)^{2}}{\epsilon}\}}]\nonumber\\
&\times&[\Re{\{\Bigl(\frac{\epsilon+2}{3}\Bigr)
^{2}\sqrt{\epsilon}-\frac{4}{9}\frac{(\epsilon-1)^{2}}{\epsilon}\}}\nonumber\\
&+&\Bigl(\frac{1}{(k_{0}R)^{3}}+\frac{1}{k_{0}R}\Bigr)
\Im{\{(\epsilon-1)\frac{\epsilon+2}{3\epsilon}\}}]\Bigr].
\end{eqnarray}
Eqs.(\ref{kres1}-\ref{Gamores}) can be compared with the results of \cite{deVriesPRL,Fleisch}.\\
\indent In summary, following \cite{MeI}, we have found analytical
expressions for the power emitted by an induced dipole
$W^{\alpha}$
--Eqs.(\ref{firstyp},\ref{secyp},\ref{thirdyp},\ref{renormy})--
and for the spontaneous emission rate $\Gamma^{\alpha}$ of a
Lorentzian-type emitter --Eqs.(\ref{Gamares},\ref{Gamores})-- from
a real cavity in a continuous medium. In the former case we
compute both the propagating and non-propagating power emission
--Eq.(\ref{n2}) and Eqs.(\ref{abs1}-\ref{abs2}) respectively-- in
function of the complex index of refraction. In the latter case,
we also compute the shift in the resonance frequency $k_{res}$
--Eqs.(\ref{kres},\ref{kres1},\ref{kres2}).
 We found that, in the small cavity limit, $\tilde{k}R\ll1$, the tuning of $k_{res}$ may be sensitively affected
by a strongly absorptive metallic environment --Eq.(\ref{kres2}). Beyond that limit and/or the continuous medium
approximation, we argue on the possibility of obtaining additional resonance frequencies.\\
%
\indent This work has been supported by Microseres-CM  and the EU
Integrated Project "Molecular Imaging" (LSHG-CT-2003-503259).

\end{document}